\documentclass[reprint,aps,prd,amsmath,eqsecnum,showpacs,nofootinbib]{revtex4-1}
\usepackage{bm,amsfonts,accents}
\usepackage{graphicx}
\newcommand{\rmd}{\mathrm{d}}
\newcommand{\rme}{\mathrm{e}}
\newcommand{\rmi}{\mathrm{i}}
\newcommand{\rmc}{\mathrm{c}}
\newcommand{\sgn}{\mathrm{sgn}}
\newcommand{\rmD}{\mathrm{D}}

\begin{document}

\title{Internal time, test clocks  and singularity resolution in dust-filled quantum cosmology}

\author{Ian D. Lawrie}
\email{i.d.lawrie@leeds.ac.uk}
\affiliation{School of Physics and Astronomy, The University of Leeds, Leeds LS2 9JT, England.}

\date{\today}

\begin{abstract}
The problem of time evolution in quantum cosmology is studied in the context of a dust-filled, spatially
flat Friedmann-Robertson-Walker universe.  In this model, two versions of the commonly-adopted notion of
internal time can be implemented in the same quantization, and are found to yield contradictory views of
the same quantum state: with one choice, the big-bang singularity appears to be resolved, but with
another choice it does not. This and other considerations lead to the conclusion that the notion of internal time
as it is usually implemented has no satisfactory physical interpretation. A recently proposed variant of the
relational-time construction, using a test clock that is regarded as internal to a specific observer,
appears to provide an improved account of time evolution relative to the proper time that elapses along
the observer's worldline. This construction permits the derivation of consistent joint probability densities
for observable quantities, which can be viewed either as evolving with proper time or as describing correlations
in a timeless manner. Section \ref{intro} reveals whether the singularity is resolved or not.

\end{abstract}

\pacs{98.80.Qc, 04.20.Cv, 04.60.Ds}

\maketitle

\section{Introduction\label{intro}}
It has long been appreciated that time evolution in generally-covariant theories such as general relativity
is, on the face of it, a gauge transformation and, contrary to everyday experience, should therefore be
unobservable.  Reviews of this `problem of time' are given, for example, in \cite{ish,kuchar,anderson},
and textbook discussions may be found in \cite{rovellibk,thiemannbk}.  In recent years, detailed studies
of quantized models of cosmology (see, e.g. \cite{bojowald,ash09,ash10,ashsingh,bojowald11} for reviews) have
demanded a practical solution to this problem, and a certain notion of `internal time' introduced by
Rovelli\cite{rovtime,rovref,rovpartial} and further developed in, for example, \cite{dittrich06,dittrich07,
thiemann06,giesel10} has been quite widely adopted. The `evolving constant of the motion' construction advocated
in these papers is a particular implementation of the idea of relational time, according to which time
evolution in covariant theories can be described only relative to the values assumed by some physical quantity that
is chosen to serve as a clock. In simple cosmological models, the clock is typically a scalar field, say $\phi$,
and in the quantum theory one obtains a wavefunction $\psi(v;\varphi)$, where $v$ is a variable describing
the geometry of the model universe, as the solution of a Schr\"odinger-like equation in which the parameter
$\varphi$ plays the role of time. In this paper, we take $v$ to be the volume of a spatially compact
universe, in which case $|\psi(v;\varphi)|^2$ is interpreted as a probability density for the volume at the
time when the scalar field $\phi$ takes the value $\varphi$.

Recently \cite{idl}, we have argued that this interpretation is hard to sustain because, for the reasons
summarized in section \ref{cjprobs} below, the values assigned to $\varphi$ cannot be regarded as readings
obtained by inspection of the `clock' $\phi$.  Indeed, according to the usual rules of quantum mechanics, the
scalar field is not an observable at all, which is especially disconcerting if it constitutes the entire matter
content of the universe. In view of this difficulty, we proposed a variant of the relational-time idea, which
makes use of a small test clock, regarded as internal to some observer from whose point of view the universe
is to be described. In this alternative construction, the test clock is unobservable, and we obtain a
wavefunction\footnote{Later on, we will adjust the notation in which various wavefunctions are expressed, so as to
maintain some important distinctions, which will be made precise in due course.}
$\psi(v,\phi;\tau)$, evolving with a time parameter $\tau$ which is not a clock reading, but corresponds
classically to the geometrical proper time that elapses along the observer's worldline. This new wavefunction
yields a joint probability density, evolving in textbook fashion with proper time $\tau$, for $v$ and $\phi$,
which are now genuine observable quantities.

In this paper, we first wish to investigate whether, despite the above-mentioned difficulties, the
internal-time wavefunction $\psi(v;\varphi)$ can be interpreted as expressing a correlation between two quantities
(the volume and scalar field), both of which are observable in some suitably broadened sense. Specifically, we
ask whether $|\psi(v;\varphi)|^2$ can be regarded as a \textit{conditional} probability density for the volume,
given that the scalar field has been determined to have the value $\varphi$.  To that end, we study a simple model
of an homogeneous universe filled with pressureless matter, described, following Brown and
Kucha\v{r}\cite{brownkuchar}, by a single scalar field. The classical version of this model is introduced in
section \ref{classical}, and we find that two complementary notions of internal time can be straightforwardly
defined, using either the volume or the scalar field as a clock.  These two internal times carry over to the
quantized theory, as discussed in section \ref{quantum}, and we use them in section \ref{internal} to compute
two corresponding probability densities.  \textit{If} the conditional-probability interpretation is feasible, then
these ought to represent, on the one hand the conditional probability density for the volume given some value
of the scalar field and, on the other hand, the conditional probability for the scalar field given some
value of the volume.  We find, however, that these probabilities are inconsistent. In fact, the two notions of
internal time lead to two mutually contradictory views of time evolution: if the volume is used as internal time,
the quantum theory appears to reproduce the classical big-bang singularity, whereas, if the scalar field is
used as internal time, \textit{exactly the same quantum state} appears to describe a universe in which the
singularity is replaced by a bounce. This result serves to strengthen our conviction that the internal-time
formalism, while resulting from perfectly sound mathematical manipulations, has no satisfactory physical
interpretation.

In section \ref{testclock}, we augment the model by the addition of a test clock.  As already indicated, the
modified relational-time construction proposed in \cite{idl} leads to a joint probability density for the
volume and scalar field, both of which which are now \textit{bona fide} observables, which
evolves, according to a standard Schr\"odinger equation, with proper time $\tau$. Like the usual time parameter
in non-relativistic physics, $\tau$ is not itself observable. In principle, no set of experimental results
records the evolution of an observed quantity relative to the time parameter that appears in Newton's laws
or in Schr\"odinger's equation.  Rather, what is recorded is an observed correlation between the object of
interest and the readings of some time-keeping device, and the rationale underlying the internal-time
formalism is that taking this practical fact seriously is the key to solving the problem of time. Consequently,
it is of some interest to recover a `timeless' interpretation, by supposing that the observer to whom our
unobserved test clock is internal also possesses a time-keeping instrument that \textit{is} observed. This
issue is taken up in section \ref{timeless}.  For the model at hand, though not in general, it turns out
that the role of a time-keeping instrument can conveniently be assigned to the scalar field (but not to
the volume).  When that is done, we obtain a genuine conditional probability density for the volume, which
coincides, to a good approximation, with the function $|\psi(v;\varphi)|^2$ found from the internal-time
formalism with $\varphi$ as the internal time. In that indirect and approximate sense, the Brown-Kucha\v{r}
scalar field emerges as a preferred internal time, indicating a bouncing universe.

\section{Classical dust-filled cosmology\label{classical}}
\subsection{The model}
As described in detail by Brown and Kucha\v{r}\cite{brownkuchar} (and in earlier work by Brown\cite{brown}),
an effective hydrodynamic description of pressureless matter, or dust, is furnished by a collection of
scalar fields $\{\phi,W_k,Z^k,M\}$ $(k=1,2,3)$, with a Lagrangian density of the form
\begin{equation}
\mathcal{L}_\mathrm{D}=-{\textstyle\frac{1}{2}}\sqrt{-g}M\left(g^{\mu\nu}U_\mu U_\nu+1\right),
\end{equation}
where the 4-velocity field is $U_\mu=-\partial_\mu\phi+W_k\partial_\mu Z^k$.  When the equations of motion
are satisfied, the integral curves of $U$ can be interpreted as the worldlines of dust particles, and the
scalar field $\phi$ is linear in the geometrical proper time $t$ that elapses along these worldlines.  We
consider the case of an homogeneous, spatially flat Friedmann-Robertson-Walker universe, with
metric $g=\mathrm{diag}(-N^2,a^2,a^2,a^2)$. To be concrete, we take spatial sections of this universe
to be compact, with coordinate volume $\int\rmd^3x=1$. Homogeneity implies that spatial derivatives of $\phi$
and $Z^k$ vanish, and variation with respect to $W_k$ leads also to $\partial_0Z^k=0$. In this special case,
therefore, the dust is modeled by a single scalar field $\phi$, along with the Lagrange multiplier $M$:
\begin{equation}
\mathcal{L}_\mathrm{D}={\textstyle\frac{1}{2}}Na^3M\left[N^{-2}(\partial_s\phi)^2-1\right],
\end{equation}
where $s$ is an arbitrary time coordinate. The momentum conjugate to $\phi$ is $p_\phi=N^{-1}a^3M\partial_s\phi$,
and variation with respect to $M$ yields the second-class constraint $p_\phi^2-(a^3M)^2=0$.  Up to a sign, this
constraint is trivially solved for $M$, and we can construct the unconstrained dust Hamiltonian
\begin{equation}
H_{\mathrm{D}}=Np_\phi.
\end{equation}
Evidently, $p_\phi$ is the total energy content of the dust, and we resolve the sign ambiguity by requiring
this to be non-negative.

In the standard way, the Einstein-Hilbert action leads to a gravitational Hamiltonian, which we write in
terms of the volume $v=a^3$ and its conjugate momentum $p_v=-(12\pi G)^{-1}N^{-1}\partial_sv/v$ as
\begin{equation}
H_\mathrm{grav}=-N(6\pi G)vp_v^2=:NC_\mathrm{grav},\label{cgrav}
\end{equation}
$G$ being the usual gravitational constant. The function $C_\mathrm{grav}$ defined by this equation is the
gravitational contribution to the Hamiltonian constraint.

Taking the lapse function $N(s)$ to be a strictly positive, but otherwise arbitrary function, we introduce
the invariant proper time
\begin{equation}
t(s)=\int_0^sN(s')\rmd s'\label{propertime}
\end{equation}
and express the Hamilton equations of motion as
\begin{eqnarray}
\dot\phi&=&N^{-1}\partial H_0/\partial p_\phi = 1\label{phieom}\\
\dot{v}&=&N^{-1}\partial H_0/\partial p_v=-(12\pi G)vp_v\label{veom}\\
\dot{p}_v&=&-N^{-1}\partial H_0/\partial v=(6\pi G)p_v^2,\label{pveom}
\end{eqnarray}
where the overdot denotes differentiation with respect to $t$. The total Hamiltonian here is $H_0=H_\mathrm{grav}
+H_\mathrm{D}$ (distinguished by its subscript from the Hamiltonian of an extended model to be considered
later) and the energy $p_\phi$ is a constant of the motion. Finally, the Hamiltonian constraint
\begin{equation}
C_0:=\partial H_0/\partial N=-(6\pi G)vp_v^2+p_\phi=0\label{constraint0}
\end{equation}
reproduces the Friedmann equation.
\subsection{Dirac observables and internal time}
The equations of motion (\ref{phieom})-(\ref{pveom}) are easily solved.  Denoting by
$\bar\phi(v,p_v,\phi,p_\phi;t)$, etc. the phase-space trajectory that passes through $(v,p_v,\phi,p_\phi)$
at $t=0$, we have
\begin{eqnarray}
\bar\phi(v,p_v,\phi,p_\phi;t)&=&\phi+t\label{phibar}\\
\bar{v}(v,p_v,\phi,p_\phi;t)&=&v(1-6\pi Gp_vt)^2\label{vbar}\\
\bar{p}_v(v,p_v,\phi,p_\phi;t)&=&p_v(1-6\pi Gp_vt)^{-1}.\label{pvbar}
\end{eqnarray}
Heuristically, given initial conditions that satisfy the constraint (\ref{constraint0}), these solutions
provide a complete description of the evolution of this simple universe relative to the proper time $t$ that
elapses along the worldline of a comoving observer. Classically, this is possible because the equations of motion
were derived \textit{before} imposing the constraint.  In a quantum-mechanical treatment, this time evolution
cannot be reproduced, because the Hamiltonian $\hat{C}_0$ that is supposed to generate it vanishes when acting on a
physically allowed state. The same difficulty arises classically in a more formal treatment of the constrained
Hamiltonian dynamics.  Here, one has to recognize that, while the proper time defined in (\ref{propertime}) is
invariant under reparametrizations of the coordinate $s$ with $N'(s')=N(s)\rmd s/\rmd s'$ (the remnant, in this
symmetry-reduced model, of the general coordinate invariance of general relativity), it is \textit{not} invariant
under an arbitrary change in the undetermined lapse function $N(s)$ unless this is accompanied by a compensating
reparametrization.  From this perspective, a change in $N(s)$, and hence evolution with respect to $t$, is
a gauge transformation generated by the constraint function $C_0$.  One has the option of fixing a gauge, by
specifying once and for all a definite function $N(s)$.  Classically, at least, the content of
(\ref{phibar})-(\ref{pvbar}) is independent of the actual function chosen, and can be regarded as physically
meaningful.  But in a manifestly gauge-independent approach, which we follow here, genuine physical information
is carried only by gauge-invariant (Dirac) observables, which commute, in the sense of Poisson brackets, with
$C_0$, and are therefore constants of the motion.

A construction due to Rovelli\cite{rovtime,rovpartial} allows us to obtain a 1-parameter family of gauge-invariant
quantities---an `evolving constant of the motion'---as follows.  Denote by $t_\varphi$ the time at which
$\bar\phi$ in (\ref{phibar}) has the value $\varphi$.  Then the quantity
\begin{equation}
V(\varphi):=\bar{v}(t_\varphi)=v[1-6\pi Gp_v(\varphi-\phi)]^2,
\end{equation}
in which we suppress the dependence on the phase-space coordinates $(v,p_v,\phi,p_\phi)$, can be interpreted
classically as `the volume at the time when the scalar field has the value $\varphi$'.  It is easy to check
explicitly that $\{V(\varphi),C_0\}=0$ for each value of the parameter $\varphi$.

A parameter such as $\varphi$ is commonly referred to in the literature as a `relational', `emergent' or `internal'
time, or simply as `time', the idea being that the scalar field serves as a physical clock, and the function
$V(\varphi)$ describes the evolution of the volume with respect to the readings of this clock. In the present
case, this may seem especially apt, in view of the linear dependence (\ref{phibar}) of the scalar field on $t$.
In \cite{brownkuchar}, indeed, this idea is enshrined in the notation: these authors use $T$ for the
scalar field that we denote by $\phi$.  For reasons that will become clear later, however, we wish to maintain
a clear distinction between the geometrical proper time defined by (\ref{propertime}) and a scalar field $\phi$
whose equation of motion \textit{happens to have the solution} (\ref{phibar}).

We now wish to define a second family of Dirac observables, taking the volume, rather than the scalar field,
as an internal time. This entails solving (\ref{vbar}) for the time $t_\nu$ at which the volume has the value
$\nu$, and requires a little care with regard to signs.  First, we take the volume always to be positive, in
contrast to the loop-quantum-gravity-inspired treatment described in \cite{aps2,acs}, where the physical volume
is the absolute value of a more fundamental variable, whose sign reflects the orientation of a co-triad.  Next,
it follows from the equation of motion (\ref{pveom}) that the evolution preserves the sign of $p_v$, so the
quantity $\sigma$ that we provisionally identify as $\sigma=-\sgn(p_v)$, is a constant of the motion. Inspection
of the solutions (\ref{vbar}) and (\ref{pvbar}) then reveals that trajectories on which $p_v$ is negative
correspond to an expanding universe, starting from an initial singularity at $t=-(6\pi G|p_v|)^{-1}$, while
those on which $p_v$ is positive correspond to a contracting universe and terminate at a final singularity
at $t=(6\pi Gp_v)^{-1}$. Consequently, we can take the square root of (\ref{vbar}), with $\bar{v}=\nu$, to
obtain
\begin{equation}
\nu^{1/2}=v^{1/2}-6\pi Gv^{1/2}p_vt_\nu,
\end{equation}
the sign of the square root being determined unambiguously by the requirement that $\nu$ is an increasing function
of $t_\nu$ when $p_v$ is negative. Thus, the Dirac observable that represents `the value of the scalar field
when the volume is $\nu$' is
\begin{equation}
\Phi(\nu):=\bar\phi(t_\nu)=\phi+(6\pi Gp_v)^{-1}-(6\pi Gv^{1/2}p_v)^{-1}\nu^{1/2}.\label{phinu}
\end{equation}
Again, one may check that $\{\Phi(\nu),C_0\}=0$ for every value of $\nu$.

Finally, with a view to quantization, we define the Dirac observables
\begin{eqnarray}
V&:=&V(0)=v+12\pi Gvp_v\phi+(6\pi G)^2vp_v^2\phi^2\label{diracV}\\
Y&:=&-(12\pi G)^{-1}V'(0)=vp_v+6\pi Gvp_v^2\phi\label{diracY}\\
\Phi&:=&\Phi(0)=\phi+(6\pi Gp_v)^{-1},\label{diracPhi}
\end{eqnarray}
and we note that $C_\mathrm{grav}$, defined in (\ref{cgrav}), is a constant of the motion, and hence
also a Dirac observable. We have the Poisson-bracket relations
\begin{eqnarray}
\{V,Y\}&=&V,\qquad\{V,C_\mathrm{grav}\}=-12\pi GY,\nonumber\\
\{Y,C_\mathrm{grav}\}&=&C_\mathrm{grav},\qquad\{\Phi,C_\mathrm{grav}\}=-1\label{poisson}
\end{eqnarray}
and the evolving observables can be expressed as
\begin{eqnarray}
V(\varphi)&=&V-12\pi GY\varphi-6\pi GC_\mathrm{grav}\varphi^2\label{vofphi}\\
\Phi(\nu)&=&\Phi+\sigma(-6\pi GC_\mathrm{grav})^{-1/2}\nu^{1/2},\label{phiofnu}
\end{eqnarray}
where we now identify
\begin{equation}
\sigma:=-\sgn(v^{1/2}p_v).\label{sigmaclass}
\end{equation}
The factor $v^{1/2}$ here, which follows from (\ref{phinu}) appears inessential, but it will prove
convenient to retain it.

We now have two notions of evolution with respect to internal time, and it will be useful to identify
the generators of these evolutions.  Indeed, we easily discover from (\ref{poisson})-(\ref{phiofnu}) that
\begin{eqnarray}
\frac{\rmd V(\varphi)}{\rmd\varphi}&=&\{V(\varphi),H_\varphi\}\label{Veom}\\
\frac{\rmd\Phi(\nu)}{\rmd\nu^{1/2}}&=&\{\Phi(\nu),H_\nu\}\label{Phieom}
\end{eqnarray}
with
\begin{eqnarray}
H_\varphi&:=&C_\mathrm{grav}\label{hphi}\\
H_\nu&:=&\sigma(-C_\mathrm{grav}/6\pi G)^{1/2}.\label{hnu}
\end{eqnarray}
On any classical trajectory, $\Phi(\nu)$ is just the inverse of $V(\varphi)$, but we shall see that this
inverse relationship is not preserved in the quantum theory.
\section{Quantum dust-filled cosmology\label{quantum}}
We consider a quantization scheme of the Wheeler-de Witt type in which, in the first instance, the canonical
coordinates $(v, p_v, \phi, p_\phi)$ are promoted to operators acting in an auxiliary (or kinematical)
vector space. A convenient representation is that in which the operators $\hat{v}$ and $\hat{p}_\phi$ act
by multiplication on wavefunctions $\Psi(v,\epsilon_\rmD)$, while their conjugate variables act by
differentiation:
\begin{eqnarray}
\hat{v}\Psi(v,\epsilon_\rmD)&=&v\Psi(v,\epsilon_\rmD),\nonumber\\
\hat{p}_\phi\Psi(v,\epsilon_\rmD)&=&\epsilon_\rmD\Psi(v,\epsilon_\rmD)\label{kinops}\\
\hat{p}_v\Psi(v,\epsilon_\rmD)&=&-\rmi\hbar\partial_v\Psi(v,\epsilon_\rmD)\nonumber\\
\hat\phi\Psi(v,\epsilon_\rmD)&=&\rmi\hbar\partial_{\epsilon_\rmD}\Psi(v,\epsilon_\rmD).\nonumber
\end{eqnarray}
The notation $\epsilon_\rmD$ reflects the fact that $\hat{p}_\phi$ corresponds to the energy content
of the dust. As in \cite{idl} we follow the authors of \cite{aps2,acs} in choosing for the gravitational
constraint the operator ordering
\begin{equation}
\hat{C}_\mathrm{grav}=-6\pi G\hat{p}_v\hat{v}\hat{p}_v.\label{cgravop}
\end{equation}
Then the Hamiltonian constraint equation (\ref{constraint0}) reads
\begin{equation}
\left[(4/\lambda^2)\partial_vv\partial_v+\epsilon_\rmD\right]\Psi(v,\epsilon_\rmD)=0,\label{constraint1}
\end{equation}
where
\begin{equation}
\lambda:=\left(\frac{2}{3\pi G\hbar^2}\right)^{1/2}.
\end{equation}
With the definition
\begin{equation}
z:=\lambda\epsilon_\rmD^{1/2}v^{1/2},
\end{equation}
the general solution to this equation can be expressed as
\begin{equation}
\Psi(v,\epsilon_\rmD)=\psi_+(\epsilon_\rmD)H_+(z)+\psi_-(\epsilon_\rmD)H_-(z),
\end{equation}
where, in terms of the usual Hankel functions, we write $H_+(z)=-\rmi H^{(2)}_0(z)$ and
$H_-(z)=\rmi H_0^{(1)}(z)$.

The classical phase space consists of two disjoint regions, distinguished by the discrete variable
(\ref{sigmaclass}), containing expanding $(\sigma =1)$ and contracting $(\sigma=-1)$ trajectories. (We
ignore, for now, the hyperplane $p_v=0$, on which the volume is constant, but see the comment following
(\ref{psiunder}).) The corresponding operator is
\begin{equation}
\hat\sigma = \sgn\left(\frac{\rmi\hbar\lambda\epsilon_\rmD^{1/2}}{2}\frac{\partial}{\partial z}\right),
\end{equation}
and we see from the integral representation
\begin{equation}
H_{\pm}(z)=\frac{2}{\pi}\int_0^\infty\rmd\xi\,\exp\left(\mp\rmi z\cosh\xi\right)
\end{equation}
that $H_+(z)$ is a linear superposition of eigenfunctions of $\rmi\partial_z$ with positive eigenvalues,
while $H_-(z)$ is a superposition of eigenfunctions with negative eigenvalues. Moreover, since
$\overline{H_+(z)}=H_-(z)$, these two components of $\Psi$ are orthogonal if, for example, we choose the
inner product
\begin{equation}
(\Psi_1,\Psi_2)=\rmi\int_0^\infty\rmd\epsilon_\rmD\int_0^\infty\rmd v\overline{\Psi}_1(v,\epsilon_\rmD)
\overleftrightarrow{\partial_v}\Psi_2(v,\epsilon_\rmD).
\end{equation}
While we will not make direct use of this inner product, these consideration provide the heuristic
motivation for our actual choice of the physical Hilbert space $\mathcal{H}_\mathrm{phys}$.  Clearly, a
solution of the constraint equation is specified by a pair of functions $\psi_\pm(\epsilon_\rmD)$, and we take
$\mathcal{H}_\mathrm{phys}$ to be the direct sum of two copies of $L^2(\mathbb{R}_+,\rmd\epsilon_\rmD)$,
with the inner product
\begin{equation}
(\psi_1,\psi_2)_\mathrm{phys}=\int_0^\infty\rmd\epsilon_\rmD\left[\left\vert\psi_+(\epsilon_\rmD)\right\vert^2
+\left\vert\psi_-(\epsilon_\rmD)\right\vert^2\right].\label{innprod}
\end{equation}
Then the operator $\hat\sigma$, which is a constant of the motion, has a well-defined action in
$\mathcal{H}_\mathrm{phys}$, namely
\begin{equation}
\hat\sigma\begin{pmatrix}\psi_+\\\psi_-\end{pmatrix}=\begin{pmatrix}\psi_+\\-\psi_-\end{pmatrix}.
\end{equation}
With the exception of $H_\nu$, defined in (\ref{hnu}), the observables we need to consider act independently
in the two subspaces, and we will use the shorthand $\hat{O}\psi(\epsilon_\rmD)$ to represent an action
of the form
\begin{equation}
\hat{O}\begin{pmatrix}\psi_+(\epsilon_\rmD)\\\psi_-(\epsilon_\rmD)\end{pmatrix}=
\begin{pmatrix}\hat{O}\psi_+(\epsilon_\rmD)\\\hat{O}\psi_-(\epsilon_\rmD)\end{pmatrix}.
\end{equation}

The operators $\hat{v}$, $\hat{p}_v$ and $\hat\phi$ defined in (\ref{kinops}) do not have any well-defined
action in $\mathcal{H}_\mathrm{phys}$; that is, they do not act on a linear combination of $H_+(z)$ and
$H_-(z)$ with $z$-independent coefficients to produce another such function.  As one might expect, however,
we can convert the classical Dirac observables $V$, $Y$ and (with slightly more difficulty) $\Phi$ given
in (\ref{diracV})-(\ref{diracPhi}) into operators that do act in $\mathcal{H}_\mathrm{phys}$. For $\hat{V}$
and $\hat{Y}$ we choose the operator orderings
\begin{eqnarray}
\hat{V}&=&\hat{v}+6\pi G(\hat{v}\hat{p}_v+\hat{p}_v\hat{v})\hat\phi-6\pi G\hat{C}_\mathrm{grav}\hat\phi^2\label{Vhat}\\
\hat{Y}&:=&{\textstyle\frac{1}{2}}(\hat{v}\hat{p}_v+\hat{p}_v\hat{v})-\hat{C}_\mathrm{grav}\hat\phi\label{Yhat},
\end{eqnarray}
with $\hat{C}_\mathrm{grav}$ given by (\ref{cgravop}). These three operators have the commutation relations
$[\hat{A},\hat{B}]=\rmi\hbar\widehat{\{A,B\}}$, with the Poisson brackets shown in (\ref{poisson}), and it follows
that the Heisenberg equation of motion (\ref{Veom}) is promoted to
\begin{equation}
\rmi\hbar\frac{\rmd \hat{V}(\varphi)}{\rmd\varphi}=[\hat{V}(\varphi),\hat{H}_\varphi]\label{Vhateom}\\
\end{equation}
with $\hat{H}_\varphi=\hat{C}_\mathrm{grav}$ and $\hat{V}(\varphi)$ the operator version of (\ref{vofphi}).
With the use of Bessel's equation $(z\partial_zz\partial_z+z^2)H_\pm(z)=0$, it is straightforward to find
the action of these operators in $\mathcal{H}_{\mathrm{phys}}$:
\begin{eqnarray}
\hat{V}\psi(\epsilon_\rmD)&=&-\frac{4}{\lambda^2}\frac{\partial}{\partial\epsilon_\rmD}\epsilon_\rmD
\frac{\partial}{\partial\epsilon_\rmD}\psi(\epsilon_\rmD)\label{Vphys}\\
\hat{Y}\psi(\epsilon_\rmD)&=&\rmi\hbar\epsilon_\rmD^{1/2}\frac{\partial}{\partial\epsilon_\rmD}
\epsilon^{1/2}_\rmD\psi(\epsilon_\rmD)\\
\hat{H}_\varphi\psi(\epsilon_\rmD)&=&\hat{C}_\mathrm{grav}\psi(\epsilon_\rmD)=-\epsilon_\rmD\psi(\epsilon_\rmD).
\end{eqnarray}
With the inner product (\ref{innprod}), $\hat{V}$ and $\hat{Y}$ are clearly symmetric, and $\hat{H}_\varphi$,
which acts by multiplication, is self-adjoint.

Construction of an operator $\hat\Phi$ corresponding to (\ref{diracPhi}) is not quite straightforward,
because $\hat{p}_v=-\rmi\hbar\partial_v$ does not have a well-defined inverse. Consider, however, a wavefunction
$\Psi(v,\epsilon_\rmD)=\psi(\epsilon_\rmD)\mathcal{C}_0(z)$, where $\mathcal{C}_0$ is any Bessel function
of order 0. We define an operator $\widehat{p_v^{-1}}$ by writing
\begin{eqnarray*}
\hat{\Phi}\Psi(v,\epsilon_\rmD)&=&\rmi\hbar\frac{\partial\Psi(v,\epsilon_\rmD)}{\partial\epsilon_\rmD}
+\frac{\rmi\hbar\lambda^2}{4}\int_{v_0}^v\Psi(v',\epsilon_\rmD)\rmd v'\\
&=&\rmi\hbar\frac{\partial\psi(\epsilon_\rmD)}{\partial\epsilon_\rmD}\mathcal{C}_0(z)+
\frac{\rmi\hbar}{2\epsilon_\rmD}\psi(\epsilon_\rmD)\Delta(z),
\end{eqnarray*}
where
$$
\Delta(z)=z\partial_z\mathcal{C}_0(z)+\int_{z_0}^z\mathcal{C}_0(z')z'\rmd z',
$$
$z_0$ is a constant, and $v_0=z^2_0/\lambda^2\epsilon_\rmD$. By virtue of Bessel's equation, we have
\begin{eqnarray*}
\Delta(z)&=&z\partial_z\mathcal{C}_0(z)-\int_{z_0}^z\partial_{z'}\left[z'\partial_{z'}\mathcal{C}_0(z')
\right]\rmd z'\\
&=&z_0\mathcal{C}_0'(z_0),
\end{eqnarray*}
and this vanishes, provided that we choose $z_0$ to be a zero of $z\mathcal{C}_0'(z)$, which is at a
complex infinity in the case of the Hankel functions.  In this way, we obtain
\begin{equation}
\hat{\Phi}\psi(\epsilon_\rmD)=\rmi\hbar\partial_{\epsilon_\rmD}\psi(\epsilon_\rmD).\label{Phiphys}
\end{equation}
This operator is symmetric under the inner product (\ref{innprod}), and has the commutator
$[\hat\Phi,\hat{C}_\mathrm{grav}]=-\rmi\hbar$, in agreement with the last Poisson-bracket relation of
(\ref{poisson}). Consequently, the equation of motion (\ref{Phieom}) becomes the operator equation
\begin{equation}
\rmi\hbar\frac{\rmd\hat\Phi(\nu)}{\rmd\nu^{1/2}}=[\hat{\Phi}(\nu),\hat{H}_\nu],
\end{equation}
where
\begin{equation}
\hat{H}_\nu=\hat\sigma \hat{G}_\nu,\quad\hat{G}_\nu\psi(\epsilon_\rmD)=\left(\frac{2\epsilon_\rmD}{3\pi G}
\right)^{1/2}\psi(\epsilon_\rmD)
\end{equation}
is self-adjoint, since it acts independently by multiplication in each subspace.

\section{Interpretation of internal time\label{internal}}
\subsection{Conditional and joint probabilities\label{cjprobs}}
We argued in \cite{idl} that, while an operator such as $\hat{V}(\varphi)$ is a perfectly good Dirac
observable, the internal time parameter $\varphi$ cannot bear the interpretation that one would like to
place on it (and which indeed is placed on it, in a significant part of the literature).

At the level of the classical equations of motion (\ref{phieom})-(\ref{pveom}) and their solutions
(\ref{phibar})-(\ref{pvbar}), it seems quite feasible to say that $V(\varphi)$ is the volume `when' the scalar
field has the value $\varphi$, provided that one does not enquire too closely about the instant of time `when'
this pair of values is realized. But even classically, if one follows the systematic procedure of constructing
a reduced, physical phase space, whose points are gauge orbits in the constraint manifold, one finds that this
physical phase space is 2-dimensional. The corresponding configuration space is 1-dimensional: there do not exist
two independent physical quantities which might simultaneously be determined to have the values $\varphi$
and $V(\varphi)$.

Quantum-mechanically, the configuration space on which states in $\mathcal{H}_\mathrm{phys}$ are defined is
again 1-dimensional. The parameter $\varphi$, which labels the family of Dirac observables $\hat{V}(\varphi)$,
\textit{cannot} be construed as a value obtained by observation of a physical clock (the scalar field), because
there is no operator acting in $\mathcal{H}_\mathrm{phys}$, independent of $\hat{V}$, to represent any such
observable clock. In particular, the operator $\hat\Phi$ cannot serve serve this purpose, for two related reasons.
First, $\hat\Phi$ does not commute with $\hat{V}(\varphi)$ for any value of $\varphi$, so the rules of quantum
mechanics do not allow simultaneous measurements of the quantities represented by these two operators. Second, the
reason these operators do not commute is that they were constructed through the `evolving constant of the motion'
algorithm, and are thus quite different from the kinematical operators $\hat{v}$ and $\hat\phi$. Classically, to
interpret $\varphi$ as a value of $\Phi$ is to interpret $V(\varphi)$ as ``the volume at the time when the value
$\varphi$ is assumed by `the scalar field at the time when the volume is zero' ''.  This is, of course,
incoherent, and is made no less so by the transition to quantum mechanics. Similar remarks apply, of course to
the parameter $\nu$ that labels the family of observables $\hat\Phi(\nu)$: it \textit{cannot} be construed as
the result of a measurement of the volume.  Given that these parameters cannot be construed as values obtained
from measurements, it is hard to see that they have any physical meaning at all.

We now wish to strengthen this conclusion by considering the possibility that, notwithstanding
the arguments just given, the internal-time formalism might be construed as yielding a joint probability
distribution that represents a correlation between observable quantities, in this case the volume and the
scalar field. That is to say, we will try to extend the notion of `observables' by dropping the requirement that
they be represented by mutually commuting operators in $\mathcal{H}_\mathrm{phys}$. This is, in particular,
a timeless interpretation, in which one might decide to ignore conundrums concerning the times at which
specific values of the observables are realized.  We will show, though, that in general no such probability
distribution exists.

The idea is this.
If $\hat{V}(\varphi)$ is regarded as the Heisenberg-picture operator associated with evolution in an internal
time $\varphi$, generated by the Hamiltonian $\hat{H}_\varphi$, then we can construct the corresponding
Schr\"odinger-picture wavefunction
\begin{equation}
\widetilde{\psi}(\tilde{\nu},\varphi)=\exp(-\rmi\hat{H}_\varphi\varphi/\hbar)
\widetilde\psi(\tilde\nu,0),\label{psiover}
\end{equation}
where $\widetilde{\psi}(\tilde\nu,0)$ is a suitable transform of $\psi(\epsilon_\rmD)$ on which $\hat{V}$ (that is,
$\hat{V}(0)$) acts by multiplication: $\hat{V}\widetilde{\psi}(\tilde\nu,0)=\tilde\nu\widetilde{\psi}(\tilde\nu,0)$.
According to the usual interpretation, the object
\begin{equation}
\widetilde{\mathcal{P}}(\tilde\nu;\varphi):=\vert\widetilde{\psi}_+(\tilde\nu,\varphi)\vert^2
+\vert\widetilde{\psi}_-(\tilde\nu,\varphi)\vert^2\label{probnuatphi}
\end{equation}
is the time-dependent probability density for obtaining the value $\tilde\nu$ from a measurement of
the volume performed at `time' $\varphi$. In particular, this probability has the time-independent
normalization $\int_0^\infty\widetilde{\mathcal{P}}(\tilde\nu;\varphi)\rmd\tilde\nu=1$. According to the foregoing
discussion, the problem with this is that we have no idea what is meant by `performing the measurement at time
$\varphi$'.

The problem might be alleviated if we could reinterpret (\ref{probnuatphi}) as a \textit{conditional}
probability density $\widetilde{\mathcal{P}}(\tilde\nu|\varphi)$ for the volume, given that some other quantity
(which we hope to identify with the scalar field) has been determined to have the value $\varphi$, even though
that other quantity does not appear explicitly in our formalism. An interpretation of that kind requires the
existence of a \textit{joint} probability density $\mathcal{P}(\tilde\nu,\varphi)$, with the normalization
\begin{equation}
\int_0^\infty\rmd\tilde\nu\int_{-\infty}^\infty\rmd\varphi\,\mathcal{P}(\tilde\nu,\varphi)=1,
\end{equation}
such that
\begin{equation}
\widetilde{\mathcal{P}}(\tilde\nu|\varphi)=\frac{\mathcal{P}(\tilde\nu,\varphi)}
{\textstyle\int_0^\infty\mathcal{P}(\tilde\nu',\varphi)\rmd\tilde\nu'}.\label{pnuphi}
\end{equation}
If this joint probability density does exist, we can use it to find the conditional probability density
\begin{equation}
\undertilde{\mathcal{P}}(\varphi|\tilde\nu)=\frac{\mathcal{P}(\tilde\nu,\varphi)}
{\textstyle\int_{-\infty}^\infty\mathcal{P}(\tilde\nu,\varphi')\rmd\varphi'}\label{pphinu}
\end{equation}
for the scalar field, given that the volume has been determined to have the value $\tilde\nu$.

For the model under consideration, it happens that we already have such a probability density to hand,
namely
\begin{equation}
\undertilde{\mathcal{P}}(\undertilde{\varphi}|\nu):=\vert\undertilde{\psi}_+(\undertilde\varphi,\nu)\vert^2
+\vert\undertilde{\psi}_-(\undertilde\varphi,\nu)\vert^2,\label{probphiatnu}
\end{equation}
where $\undertilde\psi(\undertilde\varphi,\nu)$ is the wavefunction evolved in the internal-time parameter $\nu$,
and expressed in the representation where $\hat\Phi$ acts by multiplication:
$\hat\Phi\undertilde\psi(\undertilde\varphi,\nu)=\undertilde\varphi\undertilde\psi(\undertilde\varphi,\nu)$.
More precisely, \textit{if} the expression on the right of (\ref{probnuatphi}) can be interpreted as a
conditional probability, then it must be possible to interpret (\ref{probphiatnu}) in the same way. Up to
this point, we have maintained a notational distinction between quantities that \textit{prima facie} have
quite different meanings, namely $\{\nu,\varphi\}$, which are internal-time parameters, and $\{\tilde\nu,
\undertilde\varphi\}$, which are configuration-space coordinates. However, if the internal-time formalism
is to be interpreted as expressing an observable correlation between, in this case, the volume and the scalar
field, then it must be possible to use these variables interchangeably, and we will now do so, until further
notice. In particular, we would like to identify the probability density (\ref{probphiatnu}) with the one
displayed in (\ref{pphinu}).

In the light of the arguments summarized at the beginning of this section, one might expect the conditional
probability interpretation to present some difficulty. Let us, indeed, attempt to construct the required
joint probability $\mathcal{P}(\nu,\varphi)$ from the conditional probabilities $\widetilde{\mathcal{P}}
(\nu|\varphi)$ and $\undertilde{\mathcal{P}}(\varphi|\nu)$, which can be calculated once a wavefunction
is specified.  Denote by $f(\varphi)$ the denominator in (\ref{pnuphi}) and by $g(\nu)$ the denominator
in (\ref{pphinu}). Then we have
\begin{equation}
f(\varphi)\widetilde{\mathcal{P}}(\nu|\varphi)=\mathcal{P}(\nu,\varphi)
=g(\nu)\undertilde{\mathcal{P}}(\varphi|\nu).\label{jointprob}
\end{equation}
Up to constants $f_0:=f(0)$ and $g_0:=g(0)$, which are fixed by normalization, the unknown functions are
determined by
\begin{eqnarray}
f(\varphi)&=&\frac{g_0\undertilde{\mathcal{P}}(\varphi|0)}{\widetilde{\mathcal{P}}(0|\varphi)}\\
g(\nu)&=&\frac{f_0\widetilde{\mathcal{P}}(\nu|0)}{\undertilde{\mathcal{P}}(0|\nu)}.
\end{eqnarray}
We see that the joint probability density is well defined (the first and last expressions in (\ref{jointprob})
are consistent) only if
\begin{equation}
R:=\frac{f_0\undertilde{\mathcal{P}}(\varphi|\nu)\widetilde{\mathcal{P}}(\nu|0)\widetilde{\mathcal{P}}(0|\varphi)}
{g_0\widetilde{\mathcal{P}}(\nu|\varphi)\undertilde{\mathcal{P}}(0|\nu)\undertilde{\mathcal{P}}(\varphi|0)}=1.
\label{R}
\end{equation}
In the following, we will calculate the ratio $R$ for a specific state, and find that it is not equal to 1.
This counterexample is sufficient to demonstrate that $R$ is not equal to 1 in general, but we will also
argue that our example is not especially atypical.
\subsection{Inconsistent probabilities}
We begin by constructing the wavefunction $\widetilde\psi(\tilde\nu,\varphi)$ that appears in (\ref{psiover})
and (\ref{probnuatphi}). To simplify matters, we consider a state for which $\psi_-=0$, a condition that is
preserved by evolution with respect to both $\hat{H}_\varphi$ and $\hat{H}_\nu$. That is, we focus on the
sector that corresponds classically to an expanding universe, and we will drop the subscript $_+$ that
indicates this explicitly. The representation in which the volume operator $\hat{V}$ acts by multiplication
is obtained by the Hankel transform
\begin{equation}
\widetilde{\psi}(\nu,\varphi)=\frac{\lambda}{2}\int_0^\infty\rmd\epsilon_\rmD\,
J_0(\lambda\epsilon_\rmD^{1/2}\nu^{1/2})\rme^{\rmi\epsilon_\rmD\varphi/\hbar}\psi(\epsilon_\rmD),
\label{psioverofnu}
\end{equation}
where $J_0$ is the Bessel function of the first kind. Bessel's equation $(z\partial_zz\partial_z+z^2)J_0(z)=0$
implies that the operator $\hat{V}$ defined in (\ref{Vphys}) does indeed act on this function by multiplication,
if $\psi(\epsilon_\rmD,\varphi)=\rme^{\rmi\epsilon_\rmD\varphi/\hbar}\psi(\epsilon_\rmD)$ belongs to the
domain on which $\hat{V}$ is symmetric. Provided that $\psi(\epsilon_\rmD,\varphi)$ possesses an invertible
Hankel transform, it is straightforward to verify that this transform preserves the normalization,
$\int_0^\infty\vert\widetilde\psi(\nu,\varphi)\vert^2\rmd\nu=\int_0^\infty\vert\psi(\epsilon_\rmD)\vert^2
\rmd\epsilon_\rmD$.

We consider normalized states of the form
\begin{equation}
\psi(\epsilon_\rmD)=\left[(2\beta)^{2n+1}/(2n)!\right]^{1/2}\epsilon_\rmD^n\rme^{-\beta\epsilon_\rmD},
\label{samplewfn}
\end{equation}
where $n$ is a positive integer and $\beta$ a real, positive constant, which satisfy both of these requirements.
Our main motivation for this choice is that the Hankel
transform can be calculated analytically.  In these states, the mean energy of the dust is $\bar\epsilon_\rmD
=(n+\frac{1}{2})\beta^{-1}$ and its dispersion is $\Delta\epsilon_\rmD=(2n+1)^{-1/2}\bar\epsilon_\rmD$, so for
large values of $n$ the energy distribution becomes quite sharply peaked. However, the analytic expressions
for $\widetilde\psi(\nu,\varphi)$ become very cumbersome for large $n$, so we focus on the case $n=1$.  In
that case, we find
\begin{equation}
\widetilde{\mathcal{P}}(\nu|\varphi)=(\lambda^2/\beta\gamma^3)(\gamma-\bar\nu+{\textstyle\frac{1}{4}}
\bar\nu^2)\rme^{-\bar{\nu}/\gamma},
\end{equation}
with
$$
\bar\nu:=\lambda^2\nu/2\beta,\qquad\gamma:=1+(\varphi/\beta\hbar)^2.
$$
It is easy to check that $\int_0^\infty\widetilde{\mathcal{P}}(\nu|\varphi)\rmd\nu=1$.

Evolution in the internal time $\nu^{1/2}$ is generated by $\hat{H}_\nu$, whose action on the $\sigma=+1$
subspace, and in the $\epsilon_\rmD$ representation is multiplication by $(2\epsilon_\rmD/3\pi G)^{1/2}
=\lambda\hbar\epsilon_\rmD^{1/2}$. The representation in which $\hat\Phi$ acts by multiplication is obtained
by Fourier transformation:
\begin{equation}
\undertilde\psi(\varphi,\nu)=(2\pi\hbar)^{-1/2}\int_0^\infty\rmd\epsilon_\rmD\rme^{\rmi\epsilon_\rmD\varphi/\hbar}
\rme^{-\rmi\lambda\epsilon_\rmD^{1/2}\nu^{1/2}}\psi(\epsilon_\rmD).\label{psiunder}
\end{equation}
Again, one may easily check that $\hat\Phi\undertilde\psi(\varphi,\nu)=\varphi\undertilde\psi(\varphi,\nu)$,
provided that $\psi(\epsilon_\rmD)=0$ when $\epsilon_\rmD$ is $0$ or $\infty$, which is the condition for
the operator (\ref{Phiphys}) to be symmetric. We note in passing that classically, when the constraint is
satisfied, $\epsilon_\rmD=0$ implies $p_v=0$. On this hypersurface, the volume is constant and the Dirac
observable (\ref{phinu}) is not well defined.  The need to restrict attention here to wavefunctions that
vanish at $\epsilon_\rmD=0$ is therefore not surprising.

Using the wavefunction (\ref{samplewfn}) with $n=1$, we obtain
\begin{equation}
\undertilde{\mathcal{P}}(\varphi|\nu)=\frac{2\beta^3}{\pi\hbar}\frac{\vert\chi(\rho)\vert^2}
{(\beta^2+\varphi^2/\hbar^2)^2},
\end{equation}
where $\rho:=\lambda^2\nu/(\beta-\rmi\varphi/\hbar)$ and
\begin{equation}
\chi(\rho):=\frac{1}{8}\left\{8-2\rho+\rmi\sqrt{\pi\rho}(\rho-6)\rme^{-\rho/4}\left[1-\mathrm{erf}
\left(\frac{\rmi\sqrt{\rho}}{2}\right)\right]\right\}.
\end{equation}
We do not know (and neither does any computer-algebra package available to us) how to compute the
integral $\int_{-\infty}^\infty\undertilde{\mathcal{P}}(\varphi|\nu)\rmd\varphi$ analytically, but
numerical evaluation yields the value $1$ for randomly selected values of $\nu$.

With these probabilities in hand, we find the ratio (\ref{R}) to be
\begin{equation}
R=\frac{\gamma\rme^{-\bar\nu}\vert\chi(\rho)\vert^2(1-\bar\nu+\frac{1}{4}\bar\nu^2)}
{\rme^{-\bar\nu/\gamma}\vert\chi(2\bar\nu)\vert^2(\gamma-\bar\nu+\frac{1}{4}\bar\nu^2)}.
\end{equation}
While $R$ is equal to 1 by construction when $\varphi=0$ or $\nu=0$, it is not equal to 1 elsewhere.
Consequently, the functions $\widetilde{\mathcal{P}}(\nu|\varphi)$ and $\undertilde{\mathcal{P}}(\varphi|\nu)$
\textit{cannot} consistently be interpreted as conditional probability densities arising from some
underlying joint probability distribution. Certainly, this conclusion is based on a single counterexample,
but it seems that our sample wavefunction has no special pathological feature, and the inconsistency is
very likely to be generic.  This will become a little clearer on examination of the kinds of evolution
that are associated with the two internal times $\varphi$ and $\nu$.
\subsection{Singularity resolution\label{singres}}
It is of considerable interest to see exactly what is implied by the two probability distributions. To
this end, we evaluate them using a wavefunction of the form (\ref{samplewfn}) with $n=4$.
The resulting analytic expressions are lengthy and unilluminating, but the somewhat more sharply peaked
energy distribution leads to probability densities whose nature is more readily apparent to the eye.
\begin{figure}
\includegraphics[scale=0.8]{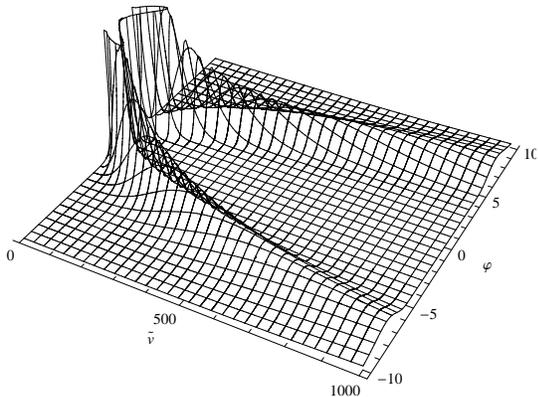}
\caption{The probability density for the volume $\tilde\nu$ evolved in the internal-time parameter $\varphi$.
The wavefunction is of the form (\ref{samplewfn}) with $n=4$, and inessential constants have the values
$\hbar=\lambda=\beta=1$.\label{fig1}}
\end{figure}

Figure \ref{fig1} shows the probability density $\widetilde{\mathcal{P}}(\tilde\nu;\varphi)$ for the volume,
evolved with the internal-time parameter $\varphi$. (We revert to the notation of (\ref{probnuatphi}), since
the conditional-probability notation has proved to be inappropriate.) Clearly, the classical singularity at
$v=0$ is resolved, in the sense that it has been replaced by a bounce in the quantum theory.  (Since the
generator $\hat{H}_\varphi$ of evolution in $\varphi$ is independent of $\hat\sigma$, the probability density
for the corresponding state in the classically contracting sector $\sigma=-1$ is exactly the same.) This agrees
qualitatively with the results of Amemiya and Koike\cite{amemiya}, who studied a similar model, adopting
the Brown-Kucha\v{r} scalar field as an internal time, though the quantization schemes they considered
differ in detail from ours. In loop-quantum-gravity-inspired treatments, such as those described in
\cite{aps1,aps2,acs}, the singularity is also seen to be resolved, but the mechanism appears to be different.
In particular, it is found in \cite{aps2} (where the matter content is a conventional massless scalar field,
which is also used as internal time) that the minimum volume at the bounce corresponds to a density
$\rho_\mathrm{crit}=3/(16\pi^2\zeta^3G^2\hbar)$, $\zeta$ being the Barbero-Immirzi parameter, independent
of the details of the quantum state.  In the present treatment, by contrast, the minimum value of
$\langle\hat{V}(\varphi)\rangle$ is just $\langle\hat{V}\rangle$, which is given by $2\beta/\lambda^2$ in
any state of the form (\ref{samplewfn}). (That the bounce occurs at $\varphi=0$ is a consequence of choosing
$\beta$ to be real in (\ref{samplewfn});  it apparent from (\ref{psioverofnu}) that an imaginary
part of $\beta$ simply shifts $\varphi$ by a constant.) This gives a maximum density $\rho_\mathrm{bounce}
=\lambda^2\bar\epsilon_\rmD^2/(2n+1)$.  Physically, a universal critical density of the order of the Planck
density seems more reasonable, so it is worth emphasizing that our purpose here is to
explore the nature of time evolution, not to construct an optimal model of cosmology, or an optimal
quantization scheme.

\begin{figure}
\includegraphics[scale=0.8]{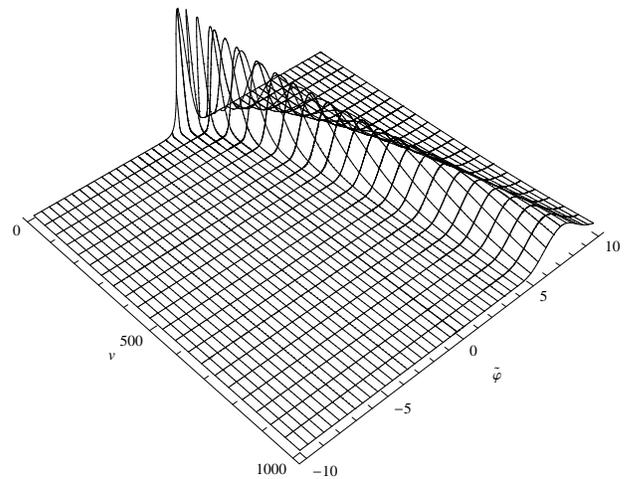}
\caption{The probability density for the scalar field $\tilde\varphi$ evolved in the internal-time
parameter $\nu$.  The quantum state is exactly the same as in FIG. \ref{fig1}.\label{fig2}}
\end{figure}

Figure \ref{fig2} shows the probability density $\undertilde{\mathcal{P}}(\tilde\varphi;\nu)$ for the
scalar field, evolved with the internal-time parameter $\nu$. It clearly indicates a universe expanding
from the initial singularity: the scalar field, which classically increases linearly with proper time,
here increases with the volume.  In the same sense, a state in the $\sigma=-1$ sector would be seen
to contract towards a final singularity. (Specifically, replacing $\epsilon_\rmD^{1/2}$ with $-\epsilon_\rmD^{1/2}$
in (\ref{psiunder}) leads to the mirror image of figure \ref{fig2}.)  From this perspective, the two sectors
of the quantum theory reproduce the expanding and contracting regions of the classical phase space.
A similar result is found in \cite{aps2} for a Wheeler-de Witt quantization of the model whose matter
content is a massless scalar field, and more recently in \cite{craig}, where the same model is treated in
a consistent-histories approach.

It is obvious from these figures that the two probability densities they depict cannot arise as
conditional probabilities from an underlying joint probability density. We emphasize that they
are computed from \textit{exactly the same quantum state}; the difference arises solely from the two
different notions of internal time used to construct the families of Dirac observables $\hat{V}(\varphi)$
and $\hat\Phi(\nu)$ and their associated Schr\"odinger-picture wavefunctions. It seems clear that these
two notions of of internal time `see' the expanding and contracting sectors of the theory very differently,
and that the difference is unlikely to be attributable to specific features of our chosen states.

\subsection{Difficulties with internal time}
The probability densities shown in figures \ref{fig1} and \ref{fig2} underline in a rather striking way
(which we did not anticipate at the outset of this investigation) the difficulties of interpretation of
the internal-time formalism discussed in section \ref{cjprobs}. It seems to us that the two notions of
internal time must stand or fall together, since they are merely two different implementations of the
same algorithm.  That is to say, since the two families of Dirac observables are constructed in the
same way, one cannot consistently maintain that $\hat{V}(\varphi)$ represents the volume when the
scalar field has the value $\varphi$ without accepting that, by the same token, $\hat\Phi(\nu)$ must
represent the scalar field when the volume is $\nu$. Since the two associated probability distributions,
which arise in \textit{exactly the same quantum state} are seen to be in gross conflict, we conclude
that both viewpoints cannot simultaneously be correct, and therefore that neither of them is correct.
As argued above, the underlying reason for this is that neither $\varphi$ nor $\nu$ can be interpreted
as a value assumed by any physically observable quantity.

Some might wish to claim that the conflict should be resolved in favour of $\varphi$ as a preferred internal
time, because the model can be `deparametrized' in this variable: that is, the constraint (\ref{constraint0})
can be solved to obtain $p_\phi = -\Theta(v,p_v)$, where the function $\Theta$ is independent of $\phi$,
and serves as the generator of evolution in the internal time $\varphi$.  We do not think that this is a
good argument in itself, but defer discussion of this point to section \ref{discussion}.

A pressing question raised by the above results is, of course,
does the quantum theory resolve the classical singularity or not?  More generally, since in this example
the answer seems to depend on an arbitrary choice of the variable that is to serve as internal time,
do we have a reliable way of deciding whether or not the singularity is resolved in the context of
\textit{any} cosmological model and quantization scheme?  A possible answer is that
quantum mechanics is actually ambivalent on this question.  It could be, for example, that given some
definite quantum state, different classes of observer will unavoidably disagree on whether that state involves a
singularity or not. We proposed in \cite{idl}, following earlier work in \cite{le}, a variant of the
idea of relational time that seems to avoid the difficulties of interpretation we have stressed up to
now, by introducing a `test clock', which provides a preferred notion of time evolution \textit{from
the point of view of a specific observer}. As described in the following section, this preferred
description, from the point of view of a comoving observer internal to the model universe, effectively
coincides with that furnished by $\varphi$ as an internal time.
\section{Dust-filled cosmology with a test clock\label{testclock}}
\subsection{Quantization of an extended model}
As explained at greater length in \cite{idl}, we supplement the model studied in previous sections
with a rough-and-ready description of a small clock, which we consider to be \textit{internal} to a
specific observer, and thus localized on that observer's worldline. In a complete description, the
coordinate functions, say $x^\mu(s)$, that locate this worldline should appear as extra phase-space
coordinates (as should a complete set of metric and matter fields, $g_{\mu\nu}(x)$ and $\phi_a(x)$),
but in the spirit of simplified cosmological models of the sort considered here, we assume that
these degrees of freedom can be neglected. If the observer is comoving, then the
proper time along the worldline is the same $t$ as appears in (\ref{propertime}). From the set of
variables $q_i$ that describe the internal workings of the clock, we suppose that a function $r(\{q\})$
can be constructed, which constitutes the reading of the clock. Crucially, however, we take the view that this
reading, being internal to the observer in question, is in principle inaccessible to that observer, and
for that reason need not feature in the physical phase space that describes the universe from that observer's
point of view. Rather, it provides the context for observations of other physical quantities to be made.
Solution of the equations of motion for the variables $\bar{q}_i(\{q\},t)$ yields the clock reading
$r(\{q\},t):=r(\{\bar{q}(\{q\},t)\})$ at the proper time $t$.  This reading need not be linear in $t$, but,
if the clock is fit for our purpose, there must be, classically, a unique function $t_0(\{q\})$, such that
$r(\{q\},t_0)=0$. That is, $t_0(\{q\})$ is the proper time at which the clock reads $0$. It is not hard to
show that $\{t_0,h\}=-1$, where $Nh$ is the clock's Hamiltonian. Because the system described by $h$ is
(ideally) localized on a single worldline, $h$ and $t_0$ are independent of the cosmological variables
$(v,p_v,\phi,p_\phi)$. The total Hamiltonian constraint is
\begin{equation}
C:=C_0+h=C_\mathrm{grav}+p_\phi+h\label{totalconstraint}
\end{equation}
and, as shown in \cite{idl}, the quantity
\begin{equation}
V(\tau):=\bar{v}(v,p_v,\phi,p_\phi;t_0+\tau)\label{Voftau}
\end{equation}
is, for each $\tau$, a Dirac observable, $\{V(\tau),C\}=0$. So, of course, are $\Phi(\tau)$, etc., defined
in the same way. Thus, we obtain a new set of evolving constants of the motion, which can be interpreted
classically as the volume, etc., when a proper time $\tau$ has elapsed along the observer's worldline
since the clock read $0$. The second crucial feature of this construction is that $\tau$ is \textit{not} to
be interpreted as a reading obtained from observation of a physical clock.  From the point of view of the
dynamical system governed by the constraint (\ref{totalconstraint}), it is an external, `heraclitian'
time, in the sense of Unruh and Wald\cite{uw}.

We quantize this enlarged model using essentially the same scheme as in section \ref{quantum}, adopting
a kinematical representation in which $\hat{v}$, $\hat{p}_\phi$ and $\hat{h}$ act by multiplication on
wavefunctions $\Psi(v,\epsilon_\rmD,\epsilon_\rmc)$, where $\epsilon_\rmc$ is the energy of the clock.
The kinematical operator representing the fiducial proper time $t_0$ is then
$\hat{t}_0=-\rmi\hbar\partial/\partial\epsilon_\rmc$.  Since we are no longer using the notion of
evolution in the internal-time parameter $\nu$, no operator of interest distinguishes the two sectors
$\sigma=\pm 1$, and we will focus on a single sector, writing a solution to the constraint equation as
\begin{equation}
\Psi(v,\epsilon_\rmD,\epsilon_\rmc)=\psi(\epsilon,\epsilon_\rmD)\mathcal{C}_0(\lambda\epsilon^{1/2}
v^{1/2}),
\end{equation}
where $\epsilon:=\epsilon_\rmD+\epsilon_\rmc$ is the total energy, and $\mathcal{C}_0$ is any Bessel
function of order 0. The physical Hilbert space is now $\mathcal{H}_\mathrm{phys}=L^2(\mathbb{R}_+^2,
\rmd\epsilon_\rmD\rmd\epsilon_\rmc)$, with the inner product
\begin{eqnarray}
(\psi_1,&&\psi_2)_\mathrm{phys}\nonumber\\
=\,&&\int_0^\infty\rmd\epsilon_\rmD\int_0^\infty\rmd\epsilon_\rmc
\bar\psi_1(\epsilon_\rmD+\epsilon_\rmc,\epsilon_\rmD)\psi_2(\epsilon_\rmD+\epsilon_\rmc,\epsilon_\rmD).
\nonumber\\
=\,&&\int_0^\infty\rmd\epsilon\int_0^\epsilon\rmd\epsilon_\rmD
\bar\psi_1(\epsilon,\epsilon_\rmD)\psi_2(\epsilon,\epsilon_\rmD).\label{innerproduct}
\end{eqnarray}
The intention is that $\epsilon_\rmc$ should be small compared with $\epsilon_\rmD$ in the same sense
that the biological system responsible for an astronomer's ciracdian rhythm, say, is small compared
with the energy content of the visible universe, and we implement this by restricting attention to
\textit{states} such that $|\psi|^2$ is very small unless $\epsilon_\rmc\ll\epsilon_\rmD$. The resulting
limitations on the resolution with which $\tau$-dependent observables might be determined are examined
in some detail in \cite{idl} for a universe whose matter content is a massless scalar field, and we will
not repeat the analysis for the present model.

Acting in $\mathcal{H}_\mathrm{phys}$, we find families of Dirac observables
\begin{eqnarray}
\hat{V}(\tau)&=&\hat{V}-8(\hbar\lambda)^{-2}\hat{Y}\tau-4(\hbar\lambda)^{-2}\hat{C}_\mathrm{grav}\tau^2
\label{Vhatoftau}\\
\hat{\Phi}(\tau)&=&\hat\Phi+\tau,\label{Phihatoftau}
\end{eqnarray}
where, acting on $\psi(\epsilon,\epsilon_\rmD)$,
\begin{eqnarray}
\hat{V}&=&-(4/\lambda^2)\partial_\epsilon\epsilon\partial_\epsilon\\
\hat{Y}&=&\rmi\hbar\epsilon^{1/2}\partial_\epsilon\epsilon^{1/2}\\
\hat{C}_\mathrm{grav}&=&-\epsilon\label{newChat}\\
\hat\Phi&=&\rmi\hbar\partial_{\epsilon_\rmD}.\label{newPhihat}
\end{eqnarray}
Note, in particular, that in (\ref{newPhihat}), the derivative is with respect to $\epsilon_\rmD$ keeping
the total energy $\epsilon$ fixed.  Treating $\epsilon_\rmD$ and $\epsilon_\rmc$ as independent variables, we have
$\hat\Phi=\hat\phi+\hat{t}_0=\rmi\hbar(\partial_{\epsilon_D}-\partial_{\epsilon_\rmc})$. Consequently,
the observables $\hat{V}(\tau)$ and $\hat\Phi(\tau)$ now commute for every $\tau$ and, according to the
usual rules of quantum mechanics, can simultaneously be assigned measured values $\underline\nu$ and
$\underline{\varphi}$. A domain on which all of these operators are symmetric is provided by the boundary conditions
\begin{equation}
\psi(\epsilon,\epsilon)=\psi(\epsilon,0)=
\partial_\epsilon\psi(\epsilon,\epsilon_\rmD)\vert_{\epsilon=\epsilon_\rmD}=0.\label{domain}
\end{equation}
Of course, $\epsilon=\epsilon_\rmD$ means that the clock energy $\epsilon_\rmc$ vanishes, and we always take
$\psi(\epsilon,\epsilon_\rmD)$ to vanish fast enough at large arguments to ensure square integrability.
\subsection{Joint probability density}
Evolution of the Heisenberg-picture operators (\ref{Vhatoftau}) and (\ref{Phihatoftau}) is generated by the
self-adjoint Hamiltonian $\hat{H}_\tau=\hat{C}_\mathrm{grav}+\hat{p}_\phi=\epsilon_\rmD-\epsilon$, and states
satisfying the boundary conditions (\ref{domain}) can be expressed in a representation in which both $\hat{V}$
and $\hat\Phi$ act by multiplication, through combined Fourier and Hankel transformation.  Thus, we can now
define a \textit{bona fide} joint probability density, which evolves in the usual way with proper time $\tau$,
namely
\begin{equation}
\mathcal{P}(\underline\nu,\underline\varphi;\tau)=\vert\psi(\underline\nu,\underline\varphi;\tau)\vert^2,
\label{tcprob}
\end{equation}
where
\begin{eqnarray}
\psi(\underline\nu,\underline\varphi;\tau)&=&\frac{\lambda}{2\sqrt{2\pi\hbar}}\int_0^\infty\rmd\epsilon
\int_0^\epsilon\rmd\epsilon_\rmD\,\rme^{\rmi\epsilon_\rmD\underline\varphi/\hbar}
\rme^{\rmi(\epsilon-\epsilon_\rmD)\tau/\hbar}\nonumber\\
&&J_0(\lambda\epsilon^{1/2}\underline\nu^{1/2})
\psi(\epsilon,\epsilon_\rmD).\label{psinuphitau}
\end{eqnarray}
In particular, this probability density has the $\tau$-independent normalization $\int_0^\infty\rmd\underline\nu
\int_{-\infty}^\infty\rmd\underline\varphi\,\mathcal{P}(\underline\nu,\underline\varphi;\tau)=1$.

To make contact with the probabilities discussed in section \ref{internal}, consider a state that
factorizes as
\begin{equation}
\psi(\epsilon,\epsilon_\rmD)=\psi(\epsilon)\psi_\rmc(\epsilon_\rmc),
\end{equation}
where $\psi_\rmc$ is a wavefunction for the clock and, as above, $\epsilon_\rmc=\epsilon-\epsilon_\rmD$
is the clock's energy. (Recall, though, that the clock is \textit{not} represented by any operator independent of
$\hat{V}$ and $\hat\Phi$ acting in $\mathcal{H}_\mathrm{phys}$, and $\tau$ is \textit{not} a value obtained
from observation of the clock). The wavefunction (\ref{psinuphitau}) becomes
\begin{eqnarray}
\psi(\underline\nu,\underline\varphi;\tau)&=&\frac{\lambda}{2}\int_0^\infty\rmd\epsilon\,\rme^{\rmi\epsilon
\underline\varphi/\hbar}J_0(\lambda\epsilon^{1/2}\underline\nu^{1/2})\psi(\epsilon)\nonumber\\
&&\times\frac{1}{\sqrt{2\pi\hbar}}\int_0^\epsilon\rmd\epsilon_\rmc\,\rme^{\rmi\epsilon_\rmc(\tau-\underline
\varphi)/\hbar}\psi_\rmc(\epsilon_\rmc).\quad\label{psinuphitau2}
\end{eqnarray}
The idea of a test clock described previously implies that $\psi(\epsilon)$ should be peaked around a value
$\bar\epsilon$ of the order of the energy content of the visible universe, while $\psi_\rmc(\epsilon_\rmc)$
should be peaked around a value $\bar\epsilon_\rmc$ somewhat smaller than the mass of an astronomer. Under
these circumstances, it is an excellent approximation to extend the upper limit of the $\epsilon_\rmc$ integral
to infinity, in which case the wavefunction (\ref{psinuphitau2}) and the probability density (\ref{tcprob})
also factorize.  In fact, we have
\begin{equation}
\mathcal{P}(\underline\nu,\underline\varphi;\tau)\simeq \widetilde{\mathcal{P}}(\underline\nu|\underline\varphi)
\mathcal{P}(\underline\varphi;\tau),\label{factorized}
\end{equation}
where $\widetilde{\mathcal{P}}(\underline\nu|\underline\varphi)$ coincides with the probability density
defined in (\ref{probnuatphi}), except that we do not now need to distinguish the sectors $\sigma=\pm1$.
Here, however, the arguments $\underline\nu$ and $\underline\varphi$ genuinely stand for values obtained
from observation of the quantities represented by the commuting operators $\hat{V}$ and $\hat\Phi$, and
$\widetilde{\mathcal{P}}(\underline\nu|\underline\varphi)$ is a genuine conditional probability. The function
\begin{equation}
\mathcal{P}(\underline\varphi;\tau):=\frac{1}{2\pi\hbar}\left\vert\int_0^\infty\rmd\epsilon_\rmc\,
\rme^{\rmi\epsilon_\rmc(\tau-\underline\varphi)/\hbar}\psi_\rmc(\epsilon_\rmc)\right\vert^2
\end{equation}
is just the probability density for obtaining the value $\underline\varphi$
from a measurement of the scalar field at proper time $\tau$.  Given a suitable choice of $\psi_\rmc(\epsilon_c)$,
it will be sharply peaked on a trajectory of the form $\underline\varphi=\Phi_0+\tau$, consistent with the
solution (\ref{Phihatoftau}) of the Heisenberg equation of motion.
\subsection{Timeless interpretation\label{timeless}}
We have emphasized that the parameter $\tau$, which labels the families of Dirac observables $\hat{V}(\tau)$
and $\hat\Phi(\tau)$ is an unobservable, external time parameter.  Classically, it coincides with the
arc length of an observer's worldline, on which the fiducial event, that the unobserved clock reads 0, serves
to define an origin. Naturally, the time-dependent observations recorded by an astronomer who wishes to
test a theoretical expression such as (\ref{tcprob}) will not refer either to $\tau$ or to the internal,
unobserved clock that we have pictured, for the sake of argument, as a biological clock. Instead, they
will refer to the observed readings of some time-keeping device, which we will call (taking a cautious
view of the generosity of the relevant funding agency) a `wristwatch'.

From an operational point of view (more or less in the sense of Bridgman\cite{bridgman}) substantive physics
is contained only in correlations between measured values of observable quantities.  The idea that a solution
to the problem of time in constrained systems is to be sought in such correlations is explicitly developed in,
for example, \cite{page, hartle,gambinietal}, and is implicit in much of the recent literature.  We have
argued that the required correlations are \textit{not} directly provided by a wavefunction such as
(\ref{psiover}), but in the model considered here, it is straightforward to see that this wavefunction does
indirectly lead to to an estimate of the desired correlation, to an approximation that might be extremely
good. In principle, we would like to study correlations between observed values of the volume, the scalar
field and the astronomer's wristwatch.  Since the wristwatch is another small clock, which classically
follows, for practical purposes, the same worldline as the unobserved test clock, we could incorporate it
into our model by adding its energy $h_\mathrm{w}$ to the total constraint:
\begin{equation}
C_\mathrm{total}:=C_\mathrm{grav}+p_\phi+h_\mathrm{w}+h.
\end{equation}
This presents no technical difficulty, but since, for this particular model, $p_\phi$ and $h_\mathrm{w}$ appear
in the same way in $C_\mathrm{total}$, it is clear that we can usefully economize on clocks by deleting
$h_\mathrm{w}$ and treating the scalar field as the astronomer's timekeeping device. (Here, we profit from the
enormous simplification that results from the assumption of homogeneity. In a more general setting, we envisage
that Dirac observables depending on $\tau$ can be constructed only from fields in the observer's immediate
locality.  Technical difficulties aside, this is no significant limitation, since the local fields include, for
example, the CMBR photons entering the astronomer's telescope.)

Suppose, then, that simultaneous measurements of volume and scalar field are performed at a sequence of
proper times $\tau_i$, distributed according to some density function $\eta(\tau)$, with
$\int_{-\infty}^\infty\eta(\tau)\rmd\tau=1$.  ($\eta(\tau)$ may have support only in a subinterval of
$(-\infty,\infty)$ if, for example, the observer's worldline terminates at a singularity.) The function
\begin{equation}
\mathcal{P}(\underline\nu,\underline\varphi):=\int_{-\infty}^\infty
\mathcal{P}(\underline\nu,\underline\varphi;\tau)\eta(\tau)\rmd\tau
\end{equation}
is a correctly normalized joint probability density, which furnishes a timeless description of the correlation
between these measured quantities.\footnote{This argument takes no account of any `collapse of the wavefunction'
occasioned by performance of the measurements.  Consequently, in the spirit of the Copenhagen interpretation,
the sequence of measurements should strictly be understood as being performed on an ensemble of identically
prepared universes.  We find this deeply unsatisfactory, but the problem concerns the interpretation of
quantum mechanics, especially as applied to the universe as a whole, about which we have nothing useful to say,
rather than the interpretation of time, about which we believe we do have something useful to say. We adopt
a Copenhagen-like point of view, not because we find it convincing, but because we find rival interpretations
to be more cumbersome without being any more convincing. While recognizing that many physicists would disagree
with this, we do not think that the present paper would be enhanced by a digression on this issue.}
In the approximation that the joint probability density factorizes as in (\ref{factorized}), we find
\begin{equation}
\mathcal{P}(\underline\nu,\underline\varphi)\simeq \widetilde{\mathcal{P}}(\underline\nu|\underline\varphi)
\mathcal{P}(\underline\varphi),
\end{equation}
where $\mathcal{P}(\underline\varphi)=\int_{-\infty}^\infty\mathcal{P}(\underline\varphi;\tau)\eta(\tau)
\rmd\tau$ is the probability for finding the measurement of the scalar field on a randomly selected occasion
to have yielded the value $\varphi$.  For the state studied in section \ref{singres},
$\widetilde{\mathcal{P}}(\underline\nu|\underline\varphi)$ is precisely the function depicted in
figure \ref{fig1}, but the introduction of an unobserved test clock now allows us to interpret this
function as a \textit{bona fide} conditional probability.  In that indirect and approximate sense, we
identify $\varphi$ as a preferred internal time, and conclude that the singularity \textit{is} resolved
in this quantum theory---or at least that the observer whose unobserved clock is represented by $h$ will
discover herself to be living in a bouncing universe.

In part, this conclusion is specific to our somewhat artificial model, in which the matter content of the
universe is provided by the Brown-Kucha\v{r} scalar field.  More generally, consider a Hamiltonian
constraint of the form
\begin{equation}
C:=C_\mathrm{grav+matt}+p_\mathrm{w}+h=0.
\end{equation}
In this expression, $C_\mathrm{grav+matt}$ is the contribution of metric and matter fields, $h$ is the
Hamiltonian of an unobserved test clock, and $p_\mathrm{w}$ is the momentum conjugate to the pointer
reading $r_\mathrm{w}$ of a wristwatch following essentially the same worldline as the test clock. By
taking the Hamiltonian of the wristwatch to be just $p_\mathrm{w}$, we model a time-keeping device that
is manufactured so as to supply a linear measure of the proper time along its worldline.  Following the
same steps as before, we expect to obtain a physical Hilbert space of functions $\psi(\bm{\gamma},\epsilon,
\epsilon_\mathrm{w})$, where $\epsilon=\epsilon_\mathrm{w}+\epsilon_\rmc$ is the total energy of the
wristwatch and the unobserved clock, while $\bm{\gamma}$ collectively denotes the remaining metric and
matter variables.  Operators on $\mathcal{H}_\mathrm{phys}$ representing Dirac observables include
\begin{eqnarray}
\hat{C}_\mathrm{grav+matt}&=&-\epsilon\\
\hat{r}_\mathrm{w}&=&\rmi\hbar\frac{\partial}{\partial\epsilon_\mathrm{w}},
\end{eqnarray}
as in (\ref{newChat}) and (\ref{newPhihat}). Taking a wave function that factorizes as
$\psi(\bm{\gamma},\epsilon,\epsilon_\mathrm{w})=\psi(\bm{\gamma},\epsilon)\psi_\rmc(\epsilon_\rmc)$, we will
obtain a Schr\"{o}dinger-picture wavefunction analogous to (\ref{psinuphitau2}), of the form
\begin{eqnarray}
\psi(\underline{\bm{\gamma}},\underline{r}_\mathrm{w};\tau)&=&\int_0^\infty\rmd\epsilon\,
\rme^{\rmi\epsilon\underline{r}_\mathrm{w}/\hbar}\psi(\underline{\bm{\gamma}},\epsilon)\nonumber\\
&&\times\int_0^\epsilon
\rmd\epsilon_\rmc\,
\rme^{\rmi\epsilon_\rmc(\tau-\underline{r}_\mathrm{w})/\hbar}\psi_\rmc(\epsilon_\rmc).
\end{eqnarray}
From this, we obtain a joint probability $\mathcal{P}(\underline{\bm{\gamma}},\underline{r}_\mathrm{w};\tau)
=|\psi(\underline{\bm{\gamma}},\underline{r}_\mathrm{w};\tau)|^2$ and, if we wish, the timeless version
$\mathcal{P}(\underline{\bm{\gamma}},\underline{r}_\mathrm{w})=
\int\mathcal{P}(\underline{\bm{\gamma}},\underline{r}_\mathrm{w};\tau)\eta(\tau)\rmd\tau$, which describe
correlations between observed values of the cosmological quantities $\bm{\gamma}$ and the wristwatch
pointer $r_\mathrm{w}$. However, because the energy $\epsilon=\epsilon_\mathrm{w}+\epsilon_\rmc$ is now
the total energy of two small clocks, the wide separation of energy scales which led to the factorization
in (\ref{factorized}) is no longer present, and the cosmological field $\varphi$ is replaced by the
wristwatch pointer $r_\mathrm{w}$. Generically, therefore, we do \textit{not} automatically recover a
conditional-probability interpretation of any internal time chosen from among the cosmological variables
$\bm{\gamma}$.
\section{Discussion\label{discussion}}
We argued in \cite{idl} that the notion of internal time, especially as commonly implemented in models of
quantum cosmology, is unsatisfactory, for two reasons. First, it offers no account of the passage of time as
it is ordinarily conceived. Ordinarily, there seems to be a clear sense in which a well-constructed clock
reads `10s' seven seconds after it read `3s', and this does not appear merely to result from a conspiracy
amongst the manufacturers of time-pieces. That is, a time-keeping device can be said to work accurately (or
not), because there is a time to be kept, not merely because its readings tend to agree (or not) with those
of other devices of the same sort.  In general relativity, this notion of time is provided, for some specific
observer,  by the proper time that elapses along that observer's worldline, but it is not recovered in a
treatment that describes evolution by correlating, for example, the volume of a spatial region with the value
of a scalar field. Second, as summarized in section \ref{cjprobs}, the meaning of a parameter that serves as
an `internal time' is unclear; in particular, it cannot be construed as a value obtained from the observation
of any physical quantity that is to be regarded as a clock, because no such quantity is represented by any
operator acting in the physical Hilbert space $\mathcal{H}_\mathrm{phys}$.

The textbook interpretations of quantum mechanics are not, of course, designed to deal with dynamics generated
by a Hamiltonian constraint, and some modification might well be needed to deal with that situation.  In
particular, one should perhaps reconsider the usual statements that observable quantities are represented by
symmetric operators on $\mathcal{H}_\mathrm{phys}$, and that the values of two such observables can be
determined simultaneously only if their corresponding operators commute. Concretely, for models of the
kind considered here, it is certainly possible to compute a wavefunction $\psi(\nu,\varphi)$, which ostensibly
evolves with respect to internal time $\varphi$, and one would like to be able to interpret it as expressing
a correlation between observed values of the volume \textit{and} the scalar field.  In view of the
$\varphi$-independent normalization $\int\vert\psi(\nu,\varphi)\vert^2\rmd\nu=1$, the quantity
$\vert\psi(\nu,\varphi)\vert^2$ cannot be interpreted as a \textit{joint} probability density, but it might
be possible to regard it as a \textit{conditional} probability density for the volume, \textit{given} that
the scalar field has the value $\varphi$.  If so, then (i) there must be some underlying joint probability
density, not directly supplied by the usual interpretation of the wavefunction, and (ii) there must exist
a complementary conditional probability density for the scalar field, given some value for the volume.

To investigate this possibility, we have studied a simple model and quantization scheme, whose main virtue
is that it admits the explicit construction \textit{within the same quantization scheme} of two complementary
internal-time evolutions. We found, though, that the two complementary probability distributions, \textit{computed
for the same quantum state} are inconsistent. In fact, these two notions of internal time lead to qualitatively
different physical interpretations of the same quantum state: with the scalar field as internal time, the universe
appears to bounce (figure \ref{fig1}), whereas with the volume as internal time it appears to expand from an
initial singularity (figure \ref{fig2}).  In view of this inconsistency, we conclude that neither of these
two notions of internal time can be correct. If the operator $\hat{V}(\varphi)$ really represents the
volume when the scalar field is $\varphi$, then it must also be true that $\hat\Phi(\nu)$ is the scalar field
when the volume is $\nu$, and that cannot be so.

We now wish to conclude further that the notion of internal time, as implemented by the evolving constant of
the motion construction, is not tenable in general. To be sure, we have discovered a serious inconsistency
only in the context of one particular model and one particular quantization scheme for which the required
computations are straightforward, but if the idea were sound in general, then it ought to apply to this
example.

As described in section \ref{testclock}, a variant of the relational-time construction proposed in
\cite{le,idl} is capable of circumventing these difficulties. We augmented the model by including an
idealized description of a small test clock, which we imagine to be internal to a specific observer, in
this case a co-moving observer.  The time parameter $\tau$ that labels families of Dirac observables is,
classically, at least, the proper time that elapses along the observer's worldline.  Its value is not to be
regarded as obtained from observation of any physical clock and, from the point of view of the observer
in question, plays the same role as the \textit{external} time in textbook Newtonian or quantum mechanics.
The test clock itself we take to be unobservable in principle by the observer in question, by virtue of being
internal to that observer.  Using this construction, we could obtain \textit{bona fide} joint probabilities,
which describe genuine correlations between cosmological observables and whatever time-keeping the observer
might use in the course of recording observations.

In the case that the observer uses the cosmological scalar field as a (large) clock, the resulting
conditional probability density for the volume essentially coincides with the evolving probability
density obtained using the scalar field as an internal time.  This is a special feature of the
Brown-Kucha\v{r} field, whose contribution to the Hamiltonian constraint (\ref{constraint0}) is just
its canonical momentum.  For the same reason, this model is deparametrizable in the variable $\phi$: in
effect, by substituting $p_\phi\to-\rmi\hbar\partial_\varphi$, one converts the constraint equation into
a Schr\"odinger-like equation that governs evolution in the internal time $\varphi$. We do not think,
however, that deparametrizability is in itself sufficient to identify a preferred internal time, though
it does, of course, make the implementation straightforward.  In general, a model is deparametrizable
in a variable $\phi$ if the constraint $C(\phi,p_\phi,\bm{\omega})=0$ can be solved to obtain
$p_\phi=-\Theta(\bm{\omega})$, where $\Theta$ is independent of $\phi$, and $\bm{\omega}$ denotes the remaining
canonical variables. Again, one can substitute $p_\phi\to-\rmi\hbar\partial_\varphi$ to obtain a
Schr\"odinger-like equation, but, again, the parameter $\varphi$ cannot be interpreted as the observed value
of some physical quantity, and it is hard to see what other meaning it might have.  Moreover, this strategy
suffers from what Kucha\v{r}\cite{kuchar} calls the `Hilbert space problem'.  That is, the Schr\"odinger-like
evolution can be implemented only if the inner product on $\mathcal{H}_\mathrm{phys}$ is chosen so as to
make $\hat\Theta$ self-adjoint. The same inner product will not necessarily confer self-adjointness on the
generators of evolution with respect to other candidates for an internal time, and it seems somewhat
unreasonable that the inner product, and hence the quantum theory as a whole, should depend on this
arbitrary choice of an internal time.  In the present example, $\phi$ has a preferred status not because
of deparametrizability as such, but rather because its contribution to the constraint is linear in $p_\phi$
with a constant coefficient. In fact, it is worth noting that the canonical transformation $u=v^{1/2}$ and
$p_u=2v^{1/2}p_v$ brings
the constraint (\ref{constraint0}) to the form $C_0=-(\lambda\hbar)^{-2}p_u^2+p_\phi=0$, so that this model is
also deparametrizable in the variable $u$. (Some authors contrast a `relativistic' deparametrization in $u$,
meaning that $C_0$ is quadratic in $p_u$, with a `non-relativistic' deparametrization in $\phi$, meaning
that $C_0$ is linear in $p_\phi$.) This simple form of $C_0$, depending only on momenta, explains why
it is straightforward, in this model, to obtain the two notions of evolution with respect to internal time
(\ref{Veom})-(\ref{hnu}), with generators that can simultaneously be made self-adjoint in the quantum theory.

We believe that the idea of a test clock, internal to some specific observer, as implemented here provides
an improved notion of time evolution, but it would be at best premature to suggest that it yields a
definitive solution to the problem of time in general. Among the limitations of the proposal as we have
so far described it are the following. (1) While the construction appears to be successful in spatially
homogeneous models with a single constraint, it does not follow that a similar construction will work
in more general spacetimes. (2) In particular, we have bypassed any explicit description of the observer's
worldline by considering only a comoving observer in a Friedmann-Robertson-Walker universe. (3) The time
parameter $\tau$ survives the passage to quantum mechanics as a $c$-number parameter, but it is not obvious
that this parameter can be unambiguously described in the quantum theory as the arc length of a worldline.
(4) In the context of simple cosmological models, at least, it seems to be an inevitable consequence of
the constraint that some object to which one is inclined to attribute a real physical existence turns out
to be unobservable. We think it is plausible that a clock which is internal to some observer should turn
out to be unobservable in a description of the universe `from that observer's point of view'. However, this
plausible form of words is not directly mandated by the formalism, and some other way of understanding the
unobservability of whatever quantity is eliminated by solution of the constraint may turn out to be better
founded.  We plan to address these issues in future work.

\end{document}